\documentclass{ws-ijmpa}

\newcommand{\fNLl} {f_{\rm NL}^{\rm local}}

\newcommand{\LL}{{\mathrm{L}}}
\newcommand{\NL}{{\mathrm{NL}}}

\usepackage{url}
\usepackage{graphicx}
\usepackage{amsmath}

\begin{document}

\markboth{Armando Bernui, Marcelo J. Rebou\c{c}as \& Antonio F. F. Teixeira}
{Large-angle Non-Gaussianity in Simulated CMB Maps}

%
\catchline{}{}{}{}{}
%

\title{LARGE-ANGLE NON-GAUSSIANITY IN SIMULATED HIGH-RESOLUTION CMB MAPS}

\author{ARMANDO BERNUI}
\address{Instituto de Ci\^encias Exatas, Universidade Federal de Itajub\'a,\\
37500-903 Itajub\'a -- MG, Brazil}

\author{MARCELO J. REBOU\c{C}AS \ AND \ ANTONIO F. F. TEIXEIRA}

\address{Centro Brasileiro de Pesquisas F\'{\i}sicas
\\ Rua Dr. Xavier Sigaud 150, \ 
22290-180 \ Rio de Janeiro -- RJ, Brazil } 

\maketitle

\begin{history}
\received{Day Month Year}
\revised{Day Month Year}
\end{history}

\begin{abstract}
A detection or nondetection of primordial non-Gaussianity by using
the cosmic microwave background radiation (CMB) offers a way of
discriminating inflationary scenarios and testing alternative models
of the early universe.
This has motivated the considerable effort that has recently gone
into the study of theoretical features of primordial non-Gaussianity
and its detection in CMB data.
Among such attempts to detect non-Gaussianity, 
there is a procedure that is based upon two indicators constructed
from the skewness and kurtosis of large-angle patches of CMB maps,
which have been proposed and used to study deviation from Gaussianity in
the WMAP data (see Refs.~\refcite{Bernui-Reboucas2009} and
\refcite{Bernui-Reboucas2010}).
Simulated CMB maps equipped with realistic primordial non-Gaussianity are essential
tools to test the viability of non-Gaussian indicators in practice,  and also to
understand the effect of systematics, foregrounds and other contaminants.
In this work we extend and complement the results
Refs.~\refcite{Bernui-Reboucas2009} and \refcite{Bernui-Reboucas2010}
by performing an analysis of non-Gaussianity  of the high-angular resolution
simulated CMB temperature maps endowed with non-Gaussianity
of the local type, for which the level of non-Gaussianity is characterized
by the dimensionless parameter  $\fNLl\,$.  


\keywords{Non-Gaussianity; cosmic microwave background radiation; inflation.}
\end{abstract}

\ccode{PACS numbers: 98.80.Es, 98.70.Vc, 98.80.-k}

\section{Introduction}  \label{Intro}

The statistical properties of the temperature 
anisotropies of cosmic microwave background (CMB) radiation 
offer a powerful probe of the physics of the primordial universe.
In particular the study of non-Gaussianity of CMB is a powerful approach to
probe the origin and evolution of structures in the universe
(see, for example, the Refs.~\refcite{Komatsu-2010}--\refcite{Xingang-Chen-2010}
and references therein). 
Given the far reaching consequences of a convincing detection (or non-detection)
of primordial non-Gaussianity for our description of the physics of the early
universe, it is important to employ different statistical tools to quantify
its amount, type and the angular scale in order to have information that may be
helpful for identifying its causes.
Apart from revealing features of non-Gaussianity, different statistical estimators
can be sensitive to different systematics. 
On the other hand, since one does not expect that a single
statistical estimator can be sensitive to all possible forms of non-Gaussianity
that may be present in CMB data, it is important to study the possible deviations
from Gaussianity by using  different statistical tools to identify any
non-Gaussian signals in the CMB data.

Recent analyses of CMB data made with different statistical tools have
provided indications of either consistency or deviation from Gaussianity
(see, e.g., Ref.~\refcite{Some_non-Gauss-refs}).
In a recent paper\cite{Bernui-Reboucas2009} we have proposed two new
large-angle non-Gaussianity indicators, based on skewness
and kurtosis of large-angle patches of CMB maps, which provide measures of the
departure from Gaussianity on large angular scales.
We have used these indicators to carry out analyses of large-angle deviation
from Gaussianity in both band and foreground-reduced WMAP CMB
maps with and without a \emph{KQ75} mask.\cite{Bernui-Reboucas2009,Bernui-Reboucas2010}
We found that while non-Gaussianity of the  Q, V,  and W  masked maps are
consistent with Gaussianity, there is a strong indication of deviation from
Gaussianity 
in the K and Ka masked maps.
We have also shown that the full-sky five-year foreground-reduced internal linear
combination (ILC)\cite{ILC-5yr-Hishaw,ILC-7yr-Gold} as well as the harmonic ILC
(HILC)~\cite{HILC-Kim} and the needlet ILC (NILC)\cite{NILC-Delabrouille}
maps present a significant deviation from Gaussianity.\cite{Bernui-Reboucas2010}

Simulated CMB maps endowed with assigned primordial non-Gaussianity 
are essential tools to test the power and sensitivity of non-Gaussian indicators.
The first reported simulations of CMB temperature maps with primordial
non-Gaussianity introduced through a non-Gaussian parameter $f_{\rm NL}$
were given by the WMAP team.\cite{NG-WMAP-1}
Subsequently, Liguori et al. produced generalized algorithm that
improves the computational speed and accuracy,\cite{Liguori_2003} and
includes polarization.\cite{Liguori_2007} A set of $300$ temperature and
polarization maps with non-Gaussianities of the local type at the WMAP angular
resolution were then produced.\cite{Liguori_2007}
More recently, Elsner and Wandelt\cite{ElsnerWandelt2009} presented  new
algorithm and generated $1\,000$ high-angular resolution simulated
non-Gaussian CMB temperature and polarization maps with non-Gaussianities of
the local type, for which the level of non-Gaussianity is defined by the
dimensionless parameter $f_{\rm NL}^{\rm local}$.

In this paper, we extend and complement the investigations  of
Refs.\refcite{Bernui-Reboucas2009} and \refcite{Bernui-Reboucas2010} (see
also the related Ref.~\refcite{Bernui-Reboucas2012}), by using
their skewness and kurtosis indicators to carry out analyses of Gaussianity
of high-angular resolution simulated non-Gaussian CMB temperature maps, equipped
with non-Gaussianity of local type with different amplitude parameters
$f_{\rm NL}^{\rm local}$, and generated according to the procedure
given by Elsner and Wandelt\cite{ElsnerWandelt2009}.

The structure of the paper is as follows. In Sec.~\ref{Indicators}
we introduce our non-Gaussianity indicators. Section~\ref{Analyses}
contains the results of applying our statistical indicators to the
non-Gaussian simulated maps and our main conclusions.

\section{Indicators and Maps of Non-Gaussianity}  \label{Indicators}

The steps of a constructive way of defining our non-Gaussianity
indicators $S$ and $K$, and the associated maps (discrete functions
defined on $S^2$) from input CMB (simulated or real data) maps are
the following:\cite{Bernui-Reboucas2009,Bernui-Reboucas2010}.
\begin{romanlist}
\item[{\bf i.}]
Take a discrete finite set of points $\{j=1, \ldots ,N_{\rm c}\}$ homogeneously
distributed on the CMB celestial sphere $S^2$ as the centers of spherical
caps of a given aperture $\gamma$; and calculate for each cap $j$ the skewness
and kurtosis  given, respectively,  by
\begin{equation} \label{S_and_K_def}
S_j   \equiv  \frac{1}{N_{\rm p} \,\sigma^3_{\!j} } \sum_{i=1}^{N_{\rm p}}
\left(\, T_i\, - \overline{T_j} \,\right)^3
\quad \mbox{and} \quad
K_j   \equiv  \frac{1}{N_{\rm p} \,\sigma^4_{\!j} } \sum_{i=1}^{N_{\rm p}}
\left(\,  T_i\, - \overline{T_j} \,\right)^4 - 3 \,,
\end{equation}
where $N_{\rm p}$ is the number of pixels in the $j^{\,\rm{th}}$ cap,
$T_i$ is the temperature at the $i^{\,\rm{th}}$ pixel, $\overline{T_j}$ is
the CMB mean temperature of the $j^{\,\rm{th}}$ cap, and $\sigma$ is the
standard deviation.
Clearly, the whole set of numbers $S_j$ and $K_j$ for each $j$, obtained
through this discrete scanning of the CMB sphere,
can be viewed as a measure of non-Gaussianity in the direction of
the center of the cap centered at $(\theta_j, \phi_j)$ with aperture $\gamma$.
\item[{\bf ii.}]
Patching together the $S_j$ and $K_j$ values for all spherical cap $j$,
give discrete functions $S = S(\theta,\phi)$ and $K = K(\theta,\phi)$
defined over the celestial sphere, which can be used to measure the deviation
from Gaussianity as a function of the angular coordinates $(\theta,\phi)$.
The Mollweide projection of skewness and kurtosis
functions $S = S(\theta,\phi)$ and $K = K(\theta,\phi)$  are nothing but
skewness and kurtosis maps, hereafter referred to as  $S-$map
and $K-$map, respectively.
\end{romanlist}

Clearly, the functions $S = S(\theta,\phi)$ and $K = K(\theta,\phi)$ are
functions defined on $S^2$ and can be expanded into their spherical
harmonics to have their power spectra $S_{\ell}$ and $K_{\ell}$.
Thus, for example, for the kurtosis indicator $K = K(\theta,\phi)$ one has
\begin{equation}
K (\theta,\phi) = \sum_{\ell=0}^\infty \sum_{m=-\ell}^{\ell}
b_{\ell m} \,Y_{\ell m} (\theta,\phi) \; ,
\end{equation}
and can calculate the corresponding angular power spectrum
\begin{equation}
K_{\ell} = \frac{1}{2\ell+1} \sum_m |b_{\ell m}|^2 \; ,
\end{equation}
which can be used to quantify the amplitude (level) and
angular scale of the deviation from Gaussianity. The power
spectrum can also be used to calculate the statistical significance
of such deviation by comparison with the corresponding power
spectrum calculated from input Gaussian maps ($f_{\rm NL}^{\rm local}=0$).
Obviously, similar expressions and analyses of the statistical significance
can be made for the skewness $S = S(\theta,\phi)$.

In the next section we shall use the statistical indicators
$S = S(\theta,\phi)$ and $K = K(\theta,\phi)$ to make
analyses of non-Gaussianity of the high-angular
resolution simulated CMB temperature maps endowed with
non-Gaussianities of the local type, for which the level
of non-Gaussianity is characterized by the dimensionless
parameter  $\fNLl\,$.  

\section{Main Results and Conclusions} \label{Analyses}

{}From the previous section it is clear that in order to calculate
the skewness and kurtosis functions $S = S(\theta,\phi)$ and
$K = K(\theta,\phi)$ and the associated $S$ and $K$ maps, 
one ought to have an input CMB map. The input maps used in our
analyses  are high-angular resolution simulated CMB temperature
maps endowed with non-Gaussianities of the local type defined
by different values of the dimensionless amplitude parameter $\fNLl$.

A simulated map with a desired level of non-Gaussianity $f_{\rm NL}^{\rm local}$ is
such that the spherical harmonic coefficients are given
by\cite{ElsnerWandelt2009}
\begin{equation}
a_{\ell m} = a^\LL_{ \ \, \ell m} + \fNLl \cdot a^\NL_{\ \ \,\,\ell m}\,,
\end{equation}
where $a^\LL_{ \ \, \ell m}$ and $a^\NL_{\ \ \,\,\ell m}$ are the linear
and non-linear spherical harmonic coefficients of the simulated CMB
non-Gaussian maps generated in Ref.~\refcite{ElsnerWandelt2009}
and are available for download.%
\footnote{
{\tt http://planck.mpa-garching.mpg.de/cmb/fnl-simulations}.}%

\begin{figure*}[htb!] \vspace{-5mm}
\begin{center}
\includegraphics[width=3.4cm,height=5.4cm,angle=90]{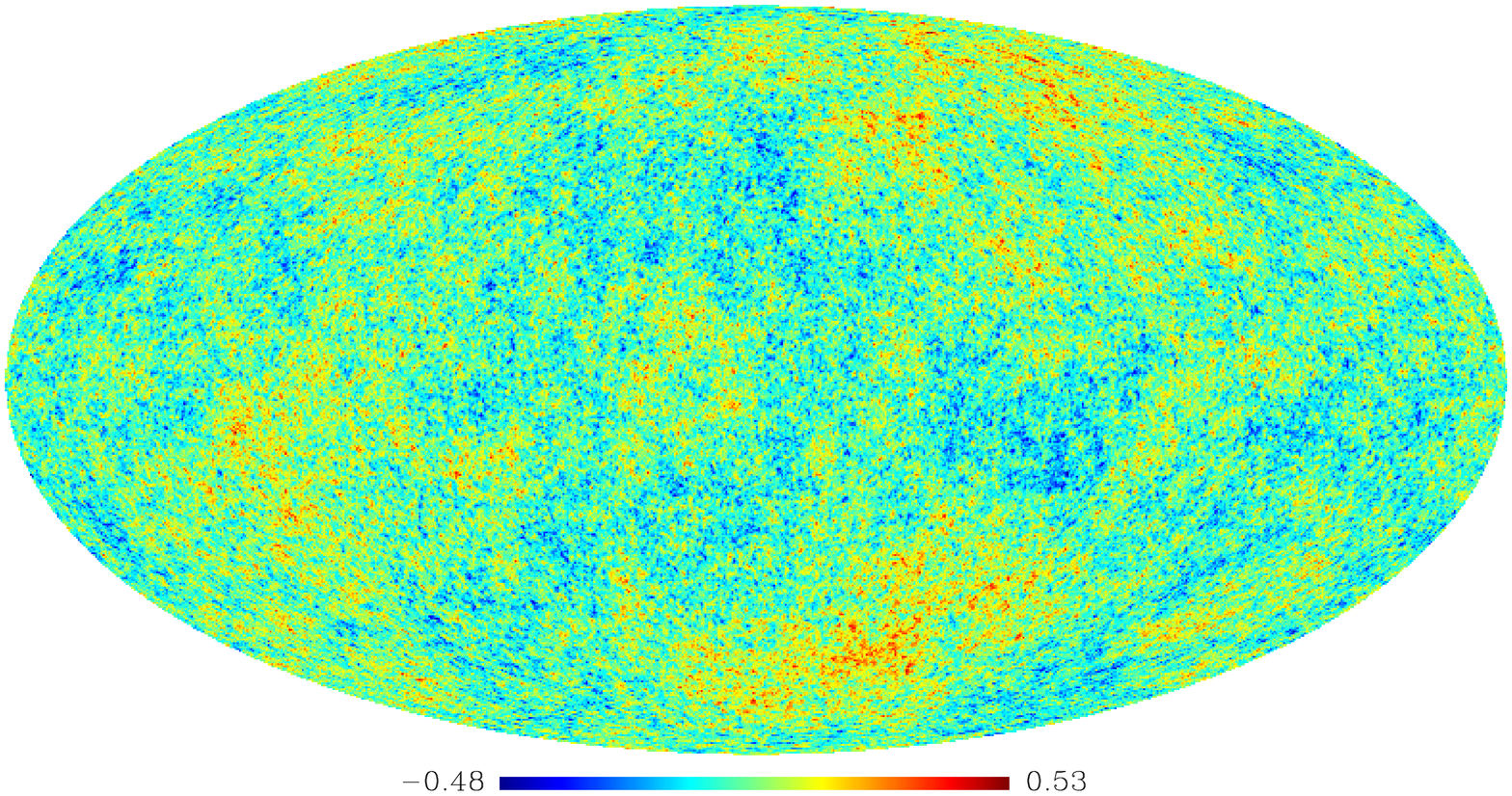} 
\hspace{0.5cm}
\includegraphics[width=3.4cm,height=5.4cm,angle=90]{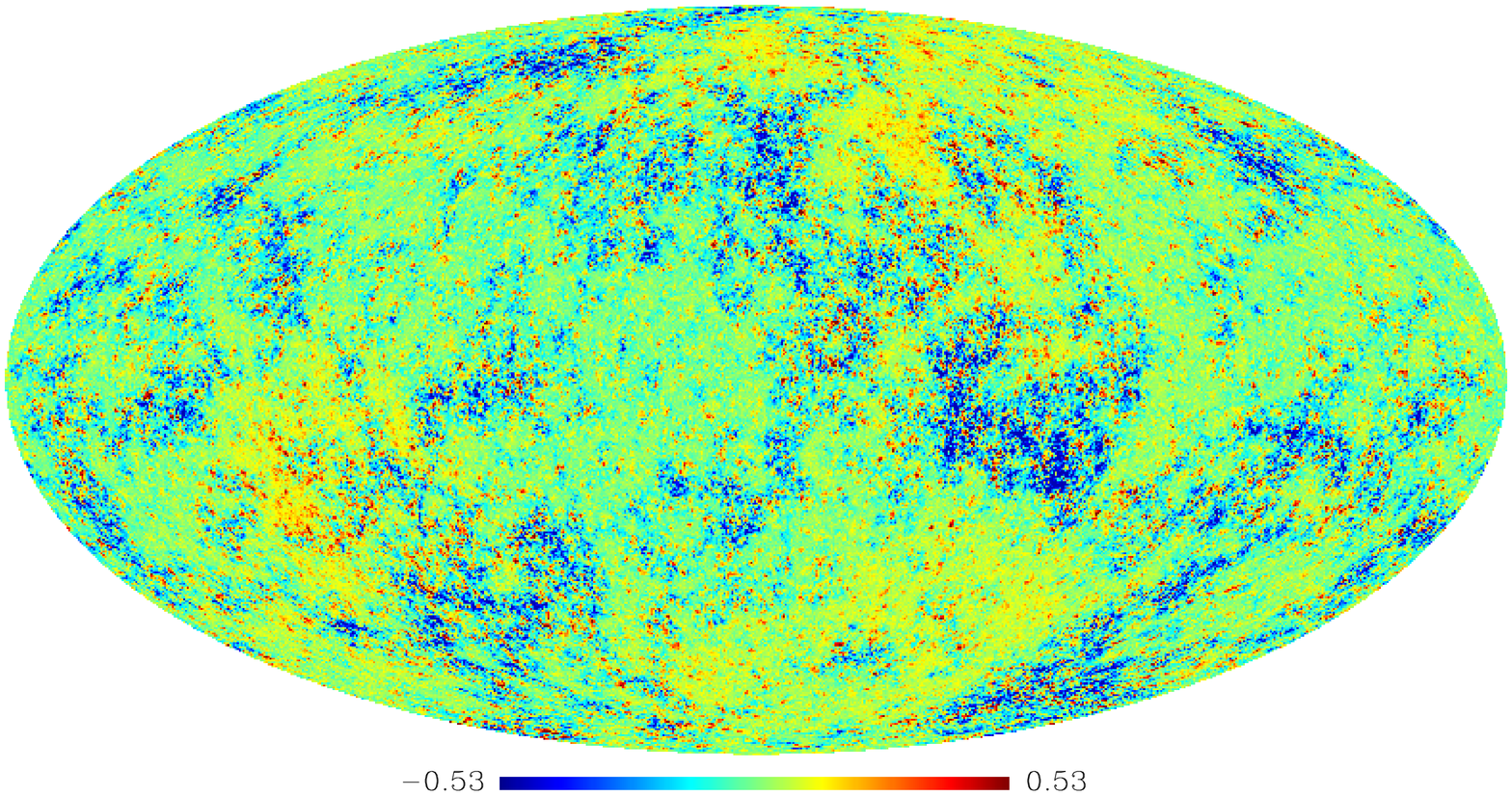} 
\caption{Simulated CMB temperature fluctuations maps. The left panel shows
a Gaussian map ($f_{\rm NL}^{\rm local}=0$) while the right panel depicts a non-Gaussian
map generated  with $f_{\rm NL}^{\rm local}=5\,000$.  We have taken this value for
$f_{\rm NL}$ just as an example that makes the non-Gaussian effects visible by naked eye
through the comparison between the Gaussian and non-Gaussian maps.
Temperatures are in $mK$. \vspace{-0.6cm}
\label{Fig1} }
\end{center}
\end{figure*}

Figure~\ref{Fig1} shows examples of such simulated input CMB maps.
The left panel gives a Gaussian map $\fNLl = 0\,$, while the right
panel shows a non-Gaussian maps with $\fNLl = 5\,000\,$.
The HEALPix resolution parameter for map of Fig.~\ref{Fig1}, as well as in
all the  high resolution simulated input CMB maps used in this paper,
is  $N_{\rm side} = 512$, which corresponds to $3\,145\,728$ pixels.
A maximum multipole moment $\ell_{\rm max} = 1\,024$ was also taken.

In the following we report the results of our analyses made
by using  $1\,000$ simulated as input CMB temperature
maps endowed with non-Gaussianities of the local type with
amplitude parameter $\fNLl = 0, 100, 1\,000, 3\,000$.

In our calculations of skewness and kurtosis indicator maps
($S-$map and $K-$map),  to minimize the statistical noise
we have scanned the celestial sphere with  spherical caps of
aperture  $\gamma = 90^{\circ}$, centered at $N_{\rm c}=3\,072$
points on the two-sphere homogeneously generated by using  HEALPix
package.\cite{Gorski-et-al-2005}

\begin{figure*}[htb!] \vspace{-5mm}
\begin{center}
\includegraphics[width=3.4cm,height=5.4cm,angle=90]{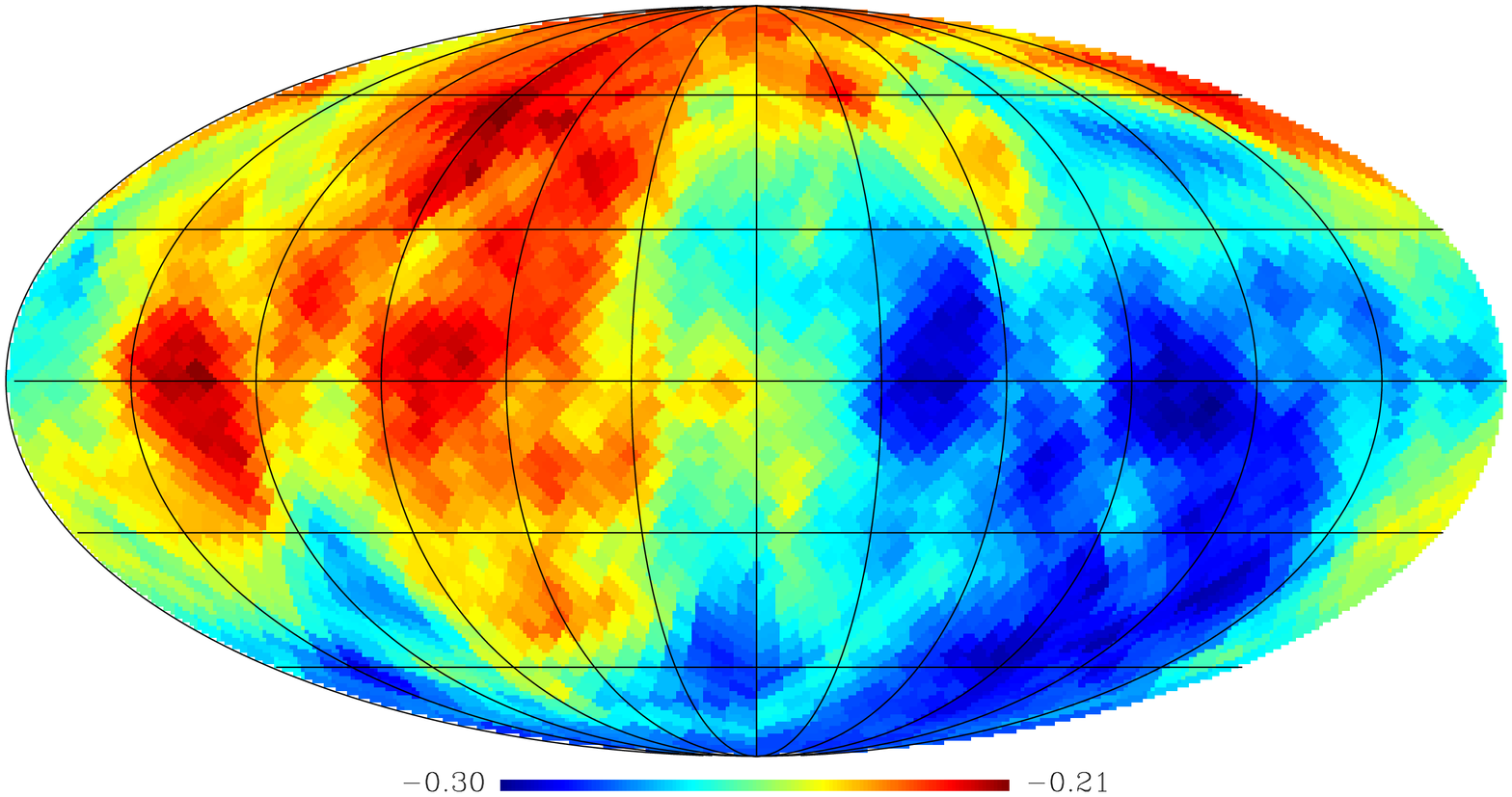} 
\hspace{0.5cm}
\includegraphics[width=3.4cm,height=5.4cm,angle=90]{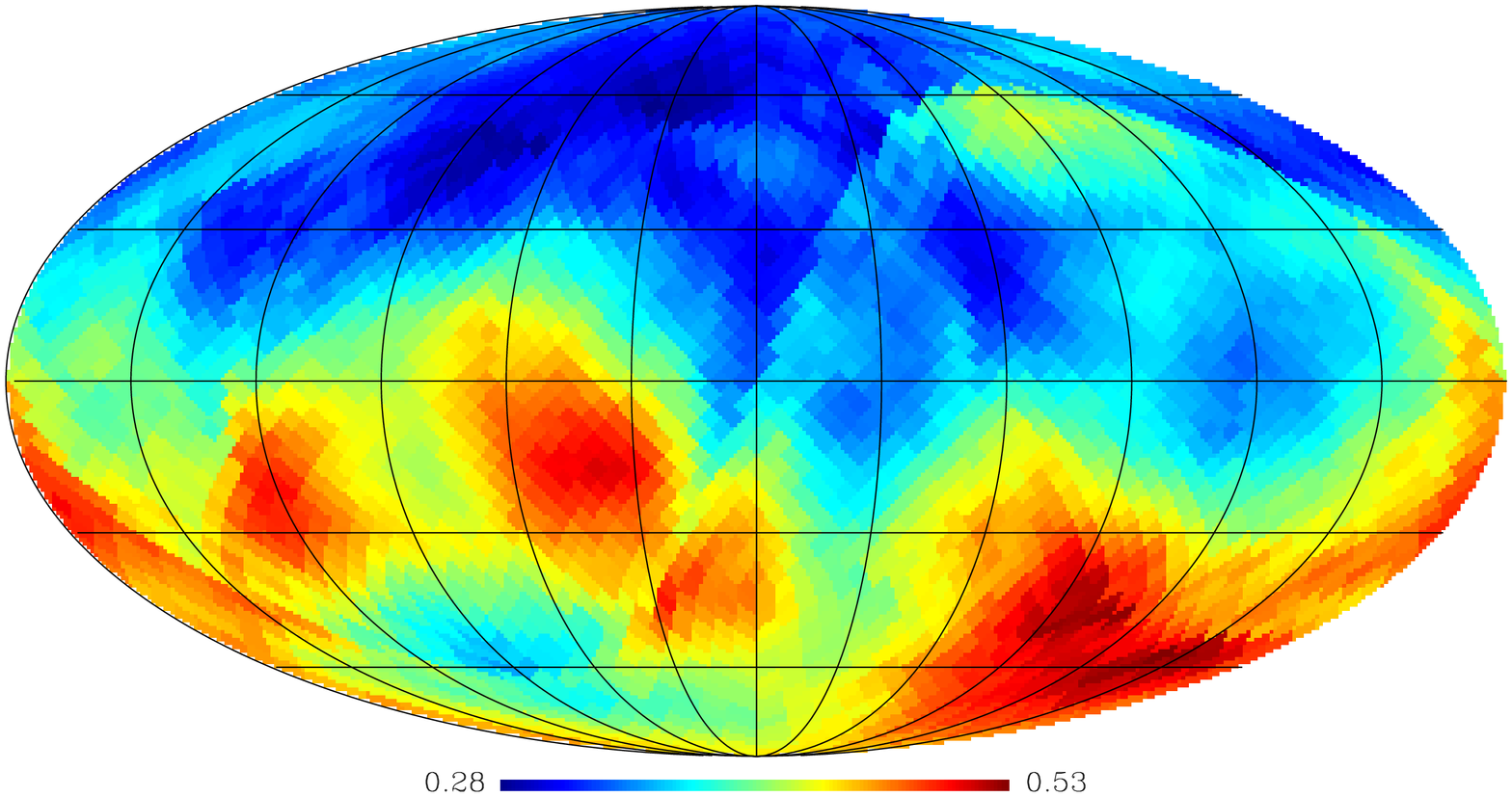} 
\caption{Skewness (left) and kurtosis (right) maps calculated from input simulated
maps with $f_{\rm NL}^{\rm local}= 1\,000$.  Clearly, according to Eq.\ref{S_and_K_def} the values
in both $S$ and $K$ are dimensionless numbers.
\vspace{-0.6cm} \label{Fig2} }
\end{center}
\end{figure*}

As an illustration of typical skewness and kurtosis maps,
Fig.~\ref{Fig2} shows the Mollweide projection of
$S$ (left) and  $K$ (right) maps generated from an input CMB simulated
map for  $f_{\rm NL}^{\rm local}=1\,000$.
These maps show  spots with higher and lower values of
$S(\theta,\phi)$ and $K(\theta,\phi)$, which suggest
\emph{large-angle}  dominant components
(low $\ell$) in these maps.
We have also calculated similar maps from the simulated input maps
endowed with non-Gaussianity of local type for the other values of
$f_{\rm NL}^{\rm local}$ that we are concerned with in this paper. However,
since these maps provide only \emph{qualitative} information, to avoid
repetition we only depict the pair of maps of Figs.~\ref{Fig2}
merely for illustrative purpose.

In order to obtain \emph{quantitative} large-angle-scale information
of the non-Gaussianity of $4\,000$ calculated $S$ and $K$ maps (obtained from
$ 4 \times 1\,000$ simulated input CMB maps generated for $f_{\rm NL}^{\rm local}= 0, 100,
1\,000, 3\,000$), we have calculated the low $\ell$ ($\,\ell=1,\,\,\cdots,10\,$)
averaged power spectra $S_{\ell}$ and $K_{\ell}$, obtained by averaging over $1\,000$
power spectra of $S$ and $K$ maps, calculated for each value of $f_{\rm NL}^{\rm local}$.
The statistical significance of these power spectra is estimated
by comparing the values of $S_{\ell}$ and $K_{\ell}$ obtained from
input maps generated for  $f_{\rm NL}^{\rm local}= 100,
1\,000, 3\,000$ with the values of the corresponding power spectra
$S_{\ell}$ and $K_{\ell}$ obtained from the Gaussian ($f_{\rm NL}^{\rm local}= 0$)
input simulated map.   

\begin{figure*}[htb!]  
\begin{center}
\includegraphics[width=6cm,height=4.3cm]{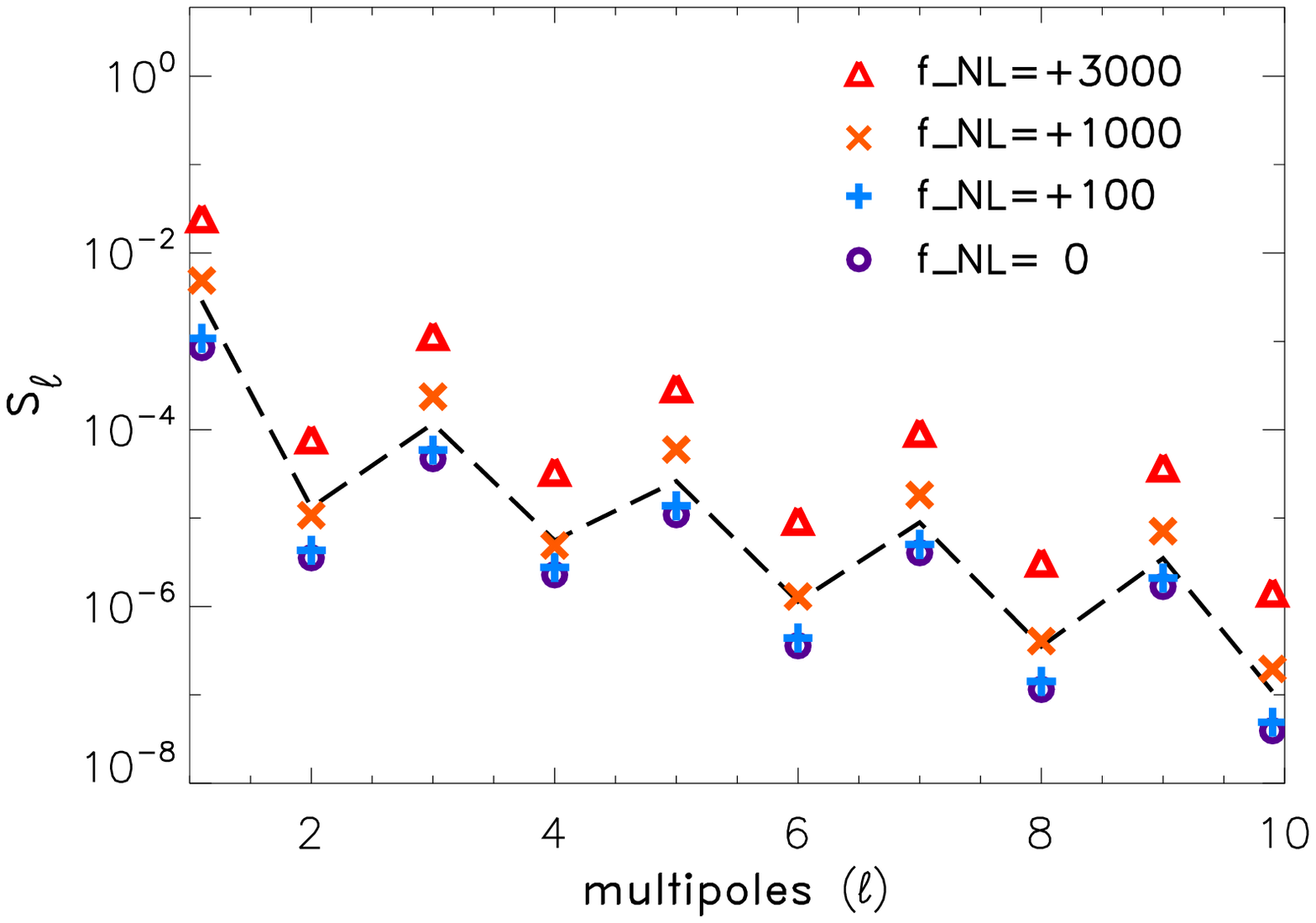}  
\hspace{0.2cm}
\includegraphics[width=6cm,height=4.3cm]{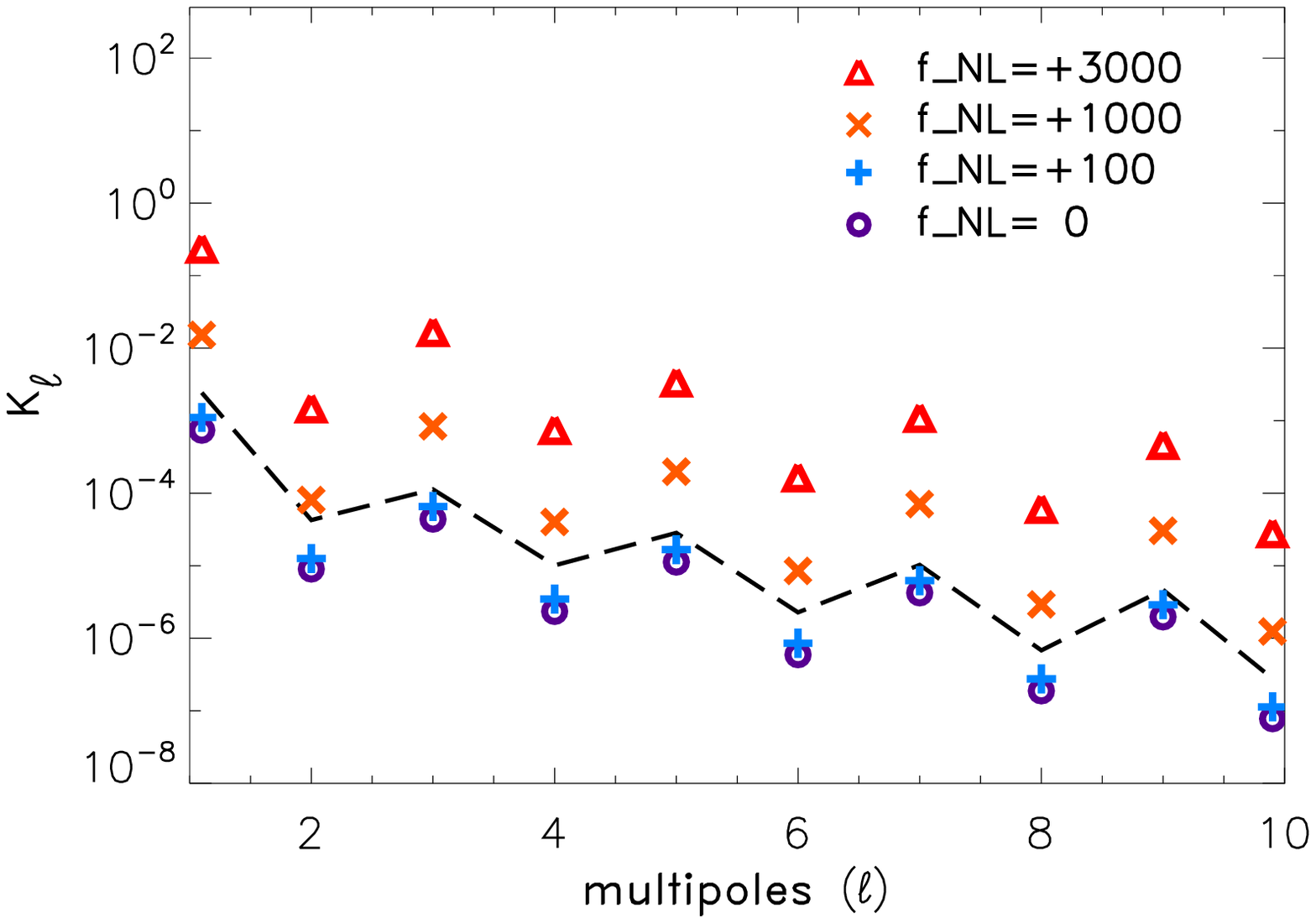}  
\caption{Low $\ell$ average power spectra of skewness $S_{\ell}$ (left) and kurtosis (right)
$K_{\ell}$ calculated from the Gaussian ($f_{\rm NL}^{\rm local}= 0$) and non-Gaussian
($f_{\rm NL}^{\rm local} \neq 0$) input simulated CMB maps. The $95\%$ confidence
level relative to the Gaussian maps is indicated by the dashed line. \vspace{-0.4cm}
\label{Fig3}  }
\end{center}
\end{figure*}

Figure~\ref{Fig3} shows the average power spectra of the skewness
$S_{\ell}$ (left panel) and kurtosis $K_{\ell}$ (right panel),
for $\,\ell=1,\,\,\cdots,10\,$, calculated from input simulated Gaussian
($f_{\rm NL}^{\rm local}= 0$) maps, and from CMB maps equipped with non-Gaussianity
of the local type for which $f_{\rm NL}^{\rm local}= 100, \,1\,000, \,3\,000$.
The $95\%$ confidence level, obtained from the $S$ and $K$ maps calculated from
the Gaussian CMB simulated maps, is indicated in this figure by the dashed line.%
\footnote{For more details on the calculation of $S$ and $K$  maps and
associated power spectra we refer the readers to Refs.~\refcite{Bernui-Reboucas2009} and
\refcite{Bernui-Reboucas2010}.}

To the extent that the average $S_{\ell}$ and $K_{\ell}$ obtained
from input simulated CMB maps endowed with $f_{\rm NL}^{\rm local}= 100$
are within $95\%$ Monte-Carlo (MC) average values of $S_{\ell}$ and $K_{\ell}$
for $f_{\rm NL}^{\rm local}= 0$, Fig.~\ref{Fig3} shows that our indicators
are not suitable to detect this small level of primordial non-Gaussianity
of local type in CMB maps. However, this figure also shows that they can
be effectively employed to detect higher level of non-Gaussianity of local type.
These results square with the refined numerical analysis we shall report
in the remainder of this paper.

To have an overall assessment power spectra $S_\ell$ and $K_\ell$,
calculated from the input simulated non-Gaussian maps equipped
primordial non-Gaussianity of local type,  we have performed a $\chi^2$
test to find out the goodness of fit for $S_{\ell}$ and $K_{\ell}$ multipole
values as compared to the expected multipole values obtained from $S$ and $K$
maps calculated from Monte-Carlo (MC) statistically Gaussian
($f_{\rm NL}^{\rm local}= 0$)  simulated CMB maps.
In each case, this gives a number that quantifies collectively the
deviation from Gaussianity.
For the power spectra $S_\ell$ and  $K_\ell$ we found the values
given in Table~\ref{table1} for the ratio $\chi^2/\text{dof}\,$
(dof stands for degrees of freedom) for the power spectra calculated
from non-Gaussianity of local type with 
$f_{\rm NL}^{\rm local}= 100, \,1\,000, \,3\,000$.

\begin{table}[th]
\tbl{$\chi^2$ test goodness of fit for $S_{\ell}$ and $K_{\ell}$
calculated from the maps with different level of non-Gaussianity
as compared with the corresponding expected values obtained
from MC simulated CMB input maps with $f_{\rm NL}^{\rm local}= 0$.}
{\begin{tabular}{@{}cll@{}} \toprule 
Level of non-Gaussianity  & \ $\chi^2$ for $S_{\ell}$  & \  $\chi^2$ for $K_{\ell}$  \\
\colrule
$f_{\rm NL}^{\rm local}= 100$    & $ 4.00 \times 10^{-3} $  & $ 3.10 \times 10^{-2} $   \\
$f_{\rm NL}^{\rm local}= 1000$   & $ 3.02 \times 10^{}   $  & $ 5.50 \times 10^2$  \\
$f_{\rm NL}^{\rm local}= 3000$   & $ 1.52 \times 10^3     $  & $ 1.98 \times 10^5 $  \\
\botrule 
\end{tabular} \label{table1}  }
\end{table}

Clearly, the greater are the values for $\chi^2/\text{dof}\,$  the smaller
are the $\chi^2$ probabilities, that is the probability that the values of
power spectra $S_{\ell}$ and $K_{\ell}$ and the expected values of the power
spectra of the Gaussian maps agree.
Thus, from  Table~\ref{table1} one concludes that the maps endowed with
$f_{\rm NL}^{\rm local}= 100$  present very small level of primordial non-Gaussianity,
as detected by our indicators, while the non-Gaussinity of the maps with
$f_{\rm NL}^{\rm local}= 1\,000 \;
\mbox{and} \; \,3\,000$ are a few orders of magnitude higher and suitably detected by
both indicators $S$ and $K$, in particular by the kurtosis indicator $K$.

\section*{Acknowledgments}

M.J. Rebou\c{c}as acknowledges the support of FAPERJ under a CNE E-26/101.556/2010 grant.
This work was also supported by Conselho Nacional de Desenvolvimento
Cient\'{\i}fico e Tecnol\'{o}gico (CNPq) - Brasil, under grant No. 475262/2010-7.
A.B. was partially supported by FAPEMIG under grant APQ-01893-10.
A.B. and M.J.R. thank CNPq for the grants under which this work
was carried out.  
We acknowledge use of the simulated maps made available
by Elsner and Wandelt\cite{ElsnerWandelt2009} and the HEALPix
package.\cite{Gorski-et-al-2005}

\end{document}